\documentclass[12pt,prl,aps,reprint,twocolumn,showpacs,floatfix,longbibliography]{revtex4-1}

\usepackage{graphicx}
\usepackage[caption=false]{subfig}
\usepackage{amsfonts}
\usepackage{amsmath}
\usepackage{mathtools}
\usepackage{float}
\usepackage{epsfig}
\usepackage{color}
\usepackage{multirow}
\usepackage[mathscr]{euscript}
\usepackage{pifont}

\newcommand{\n}{\eta}

\newcommand{\ee}[1]{\ensuremath{10^{#1}}}

\begin{document}

\title{Uncovering Multiscale Order in the Prime Numbers via Scattering}


\author{S. Torquato}

\email{Corresponding author: torquato@princeton.edu}

\affiliation{\emph{Department of Chemistry, Department of Physics,
Princeton Institute for the Science and Technology of
Materials, and Program in Applied and Computational Mathematics}, \emph{Princeton University},
Princeton NJ 08544}

\author{G. Zhang}

\affiliation{\emph{Department of Chemistry}, \emph{Princeton University},
Princeton NJ 08544 }

\author{M. de Courcy-Ireland }


\affiliation{\emph{Department of Mathematics}, \emph{Princeton University},
Princeton NJ 08544}

\begin{abstract}

The prime numbers have been a source of fascination for millenia and continue to surprise us.
Motivated by the hyperuniformity concept, which has attracted recent attention in physics
and materials science, we show that the prime numbers in certain large intervals possess unanticipated order across length scales and represent the first example of a new class of many-particle systems with pure point diffraction patterns, which we call {\it effectively limit-periodic}. In particular, the primes in this regime are hyperuniform. This is shown analytically
using the structure factor $S(k)$, proportional to the scattering intensity from a many-particle system.
Remarkably, the structure factor for primes is characterized by dense Bragg peaks, like a quasicrystal, but
positioned at certain rational wavenumbers, like a limit-periodic point pattern. We identify a transition between ordered and disordered
prime regimes that depends on the intervals studied. Our analysis leads
to an algorithm that enables one to predict primes with high accuracy.
Effective limit-periodicity deserves future investigation in physics, independent of
its link to the primes.

\end{abstract}

\maketitle

\section*{Introduction}

\begin{figure*}[bthp]
\subfloat[]{\label{fig:FIG1a}%
\includegraphics[width=0.4\textwidth,clip=]{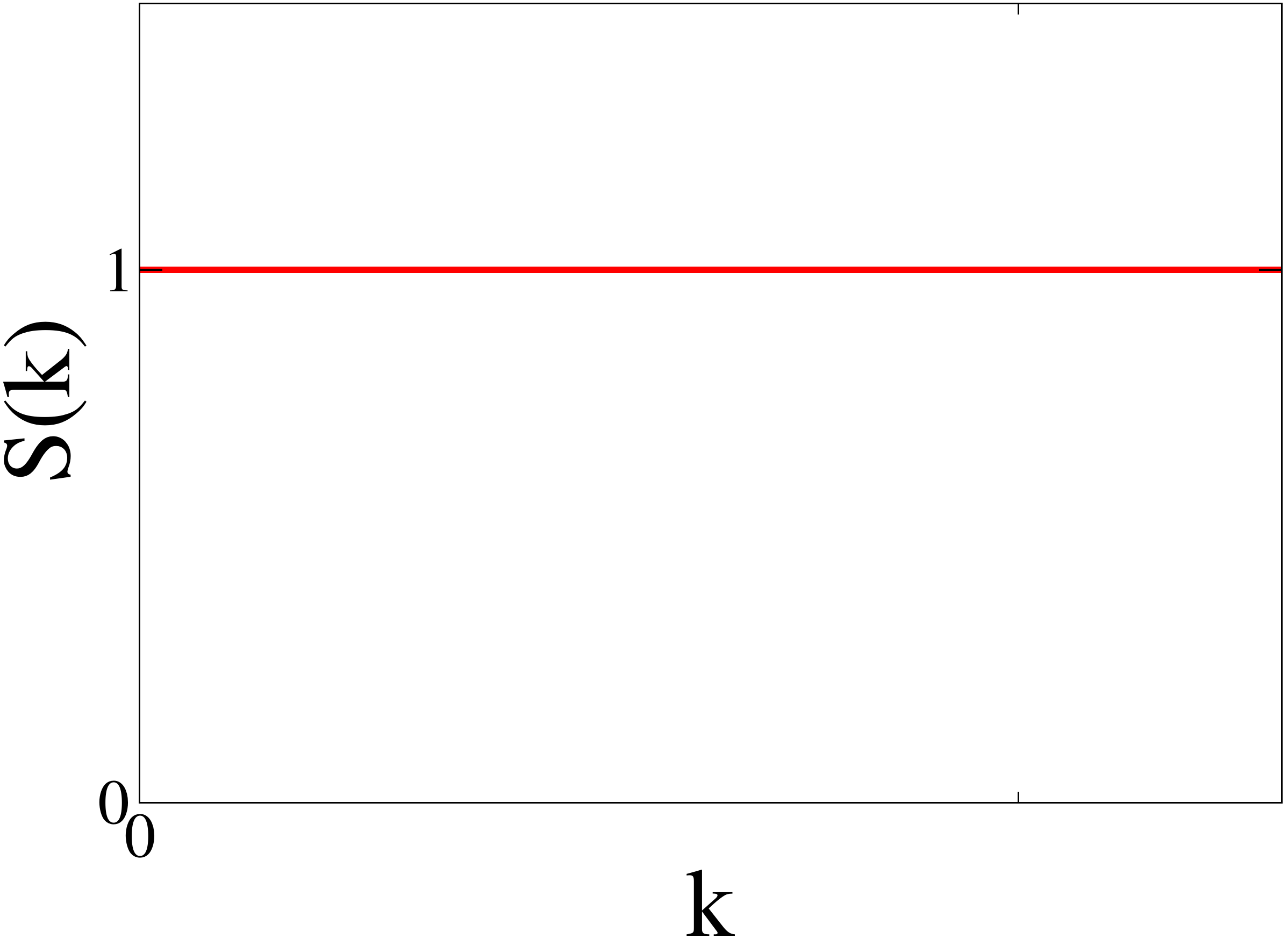}
}
\subfloat[]{\label{fig:FIG1b}%
\includegraphics[width=0.4\textwidth,clip=]{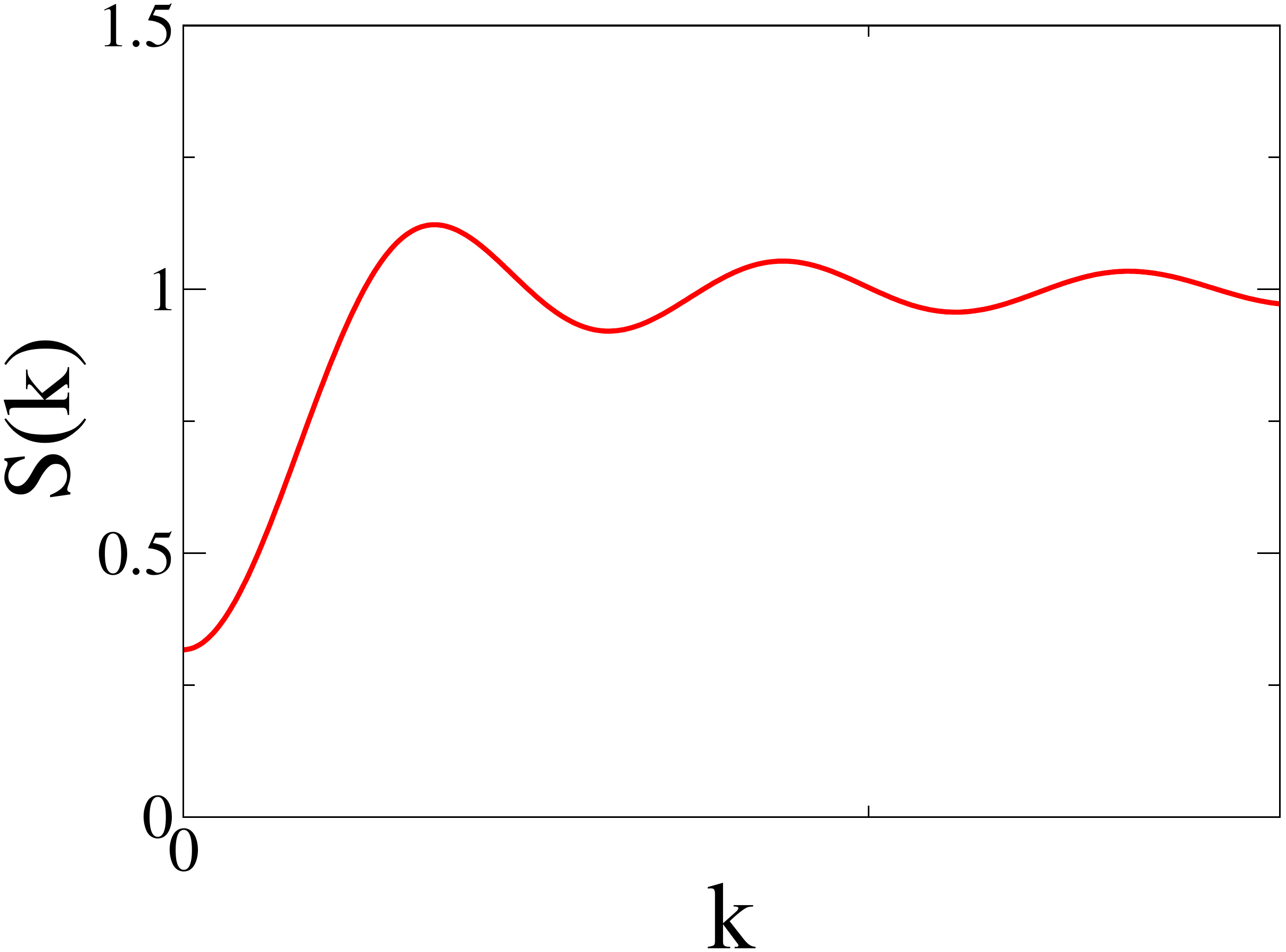} 
}\\
\subfloat[]{\label{fig:FIG1c}%
\includegraphics[width=0.4\textwidth,clip=]{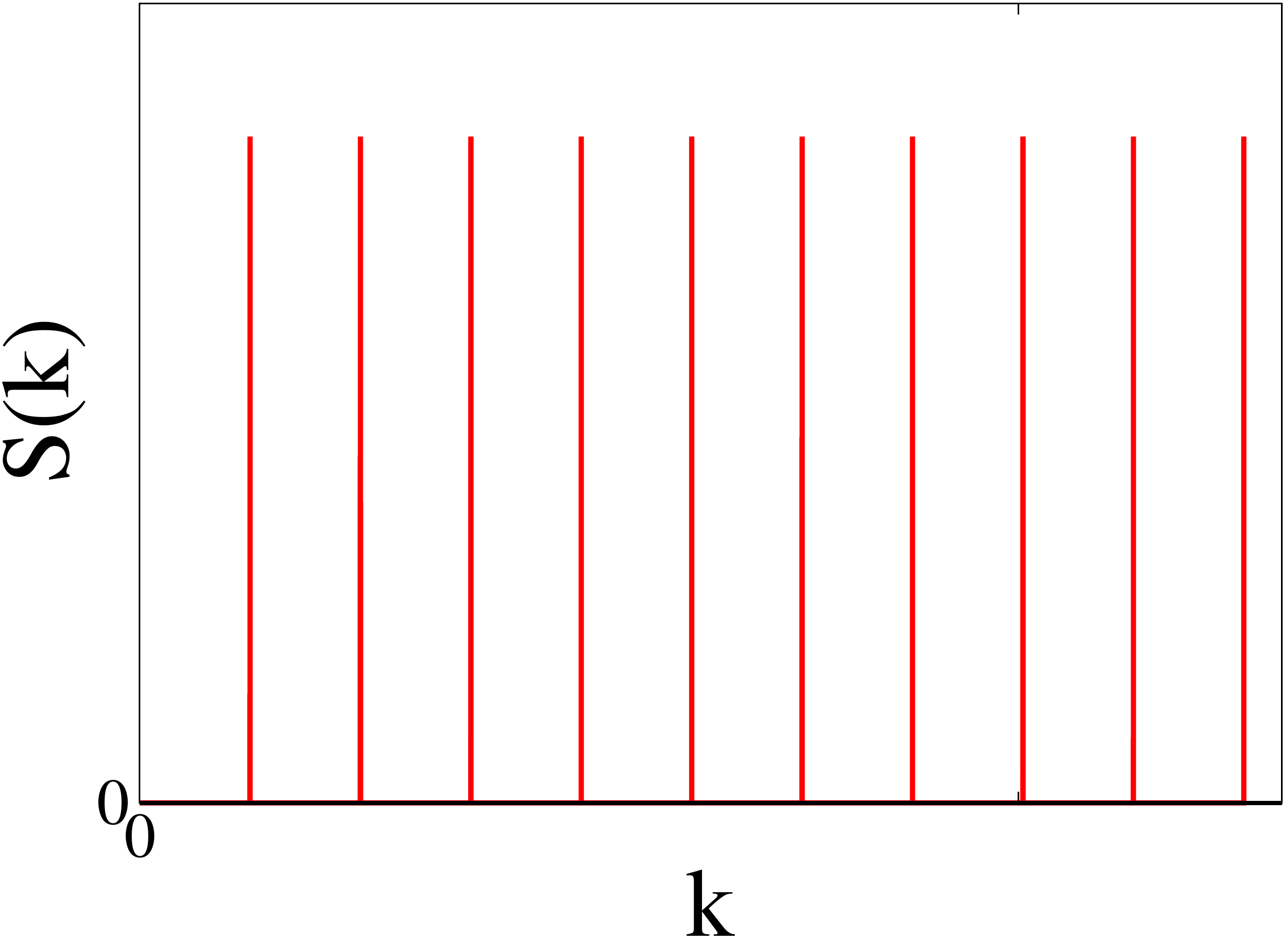}
}
\subfloat[]{\label{fig:FIG1d}%
\includegraphics[width=0.4\textwidth,clip=]{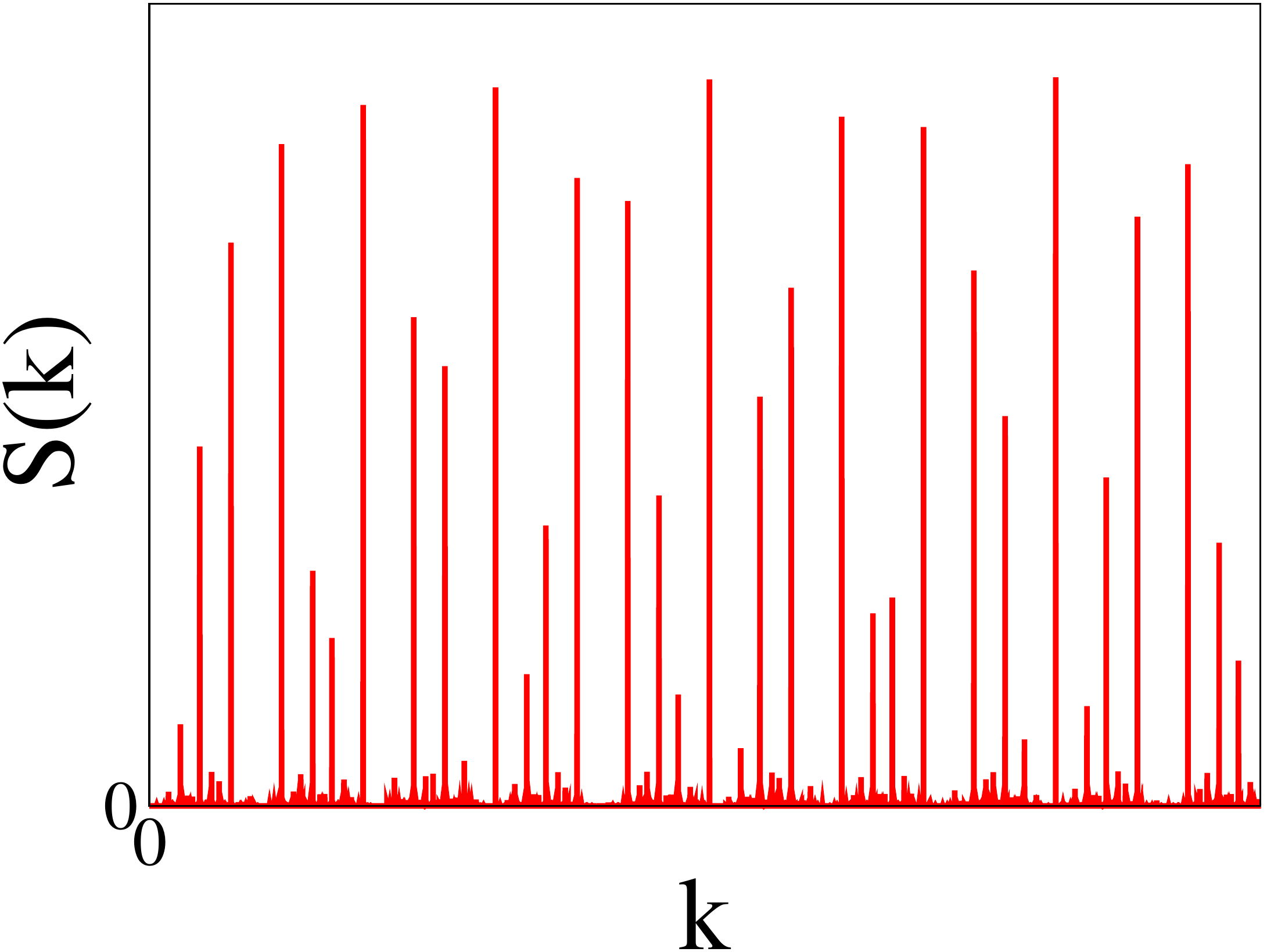}
}\\
\subfloat[]{\label{fig:FIG1e}%
\includegraphics[width=0.4\textwidth,clip=]{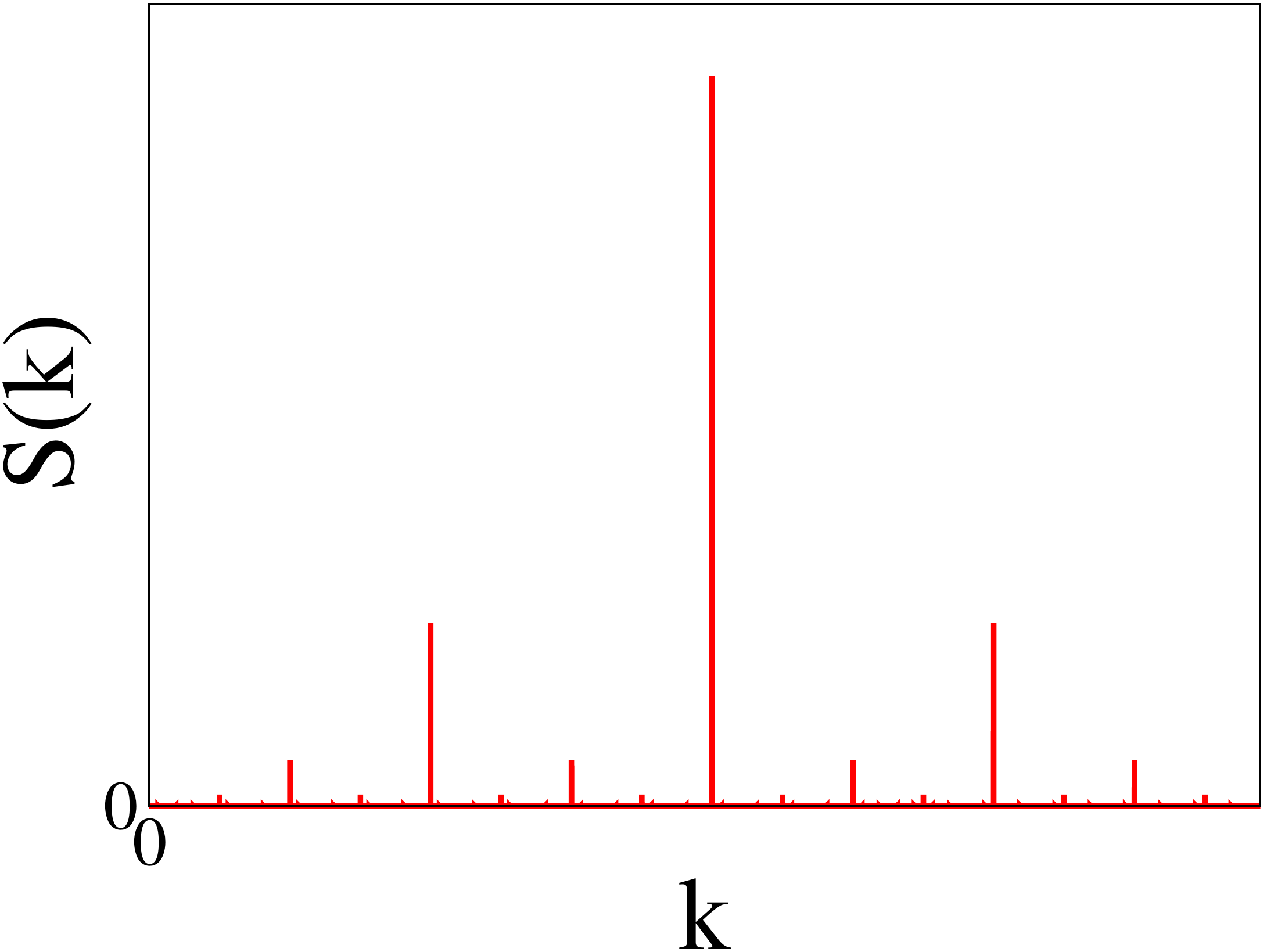} 
}
\subfloat[]{\label{fig:FIG1f}%
\includegraphics[width=0.4\textwidth,clip=]{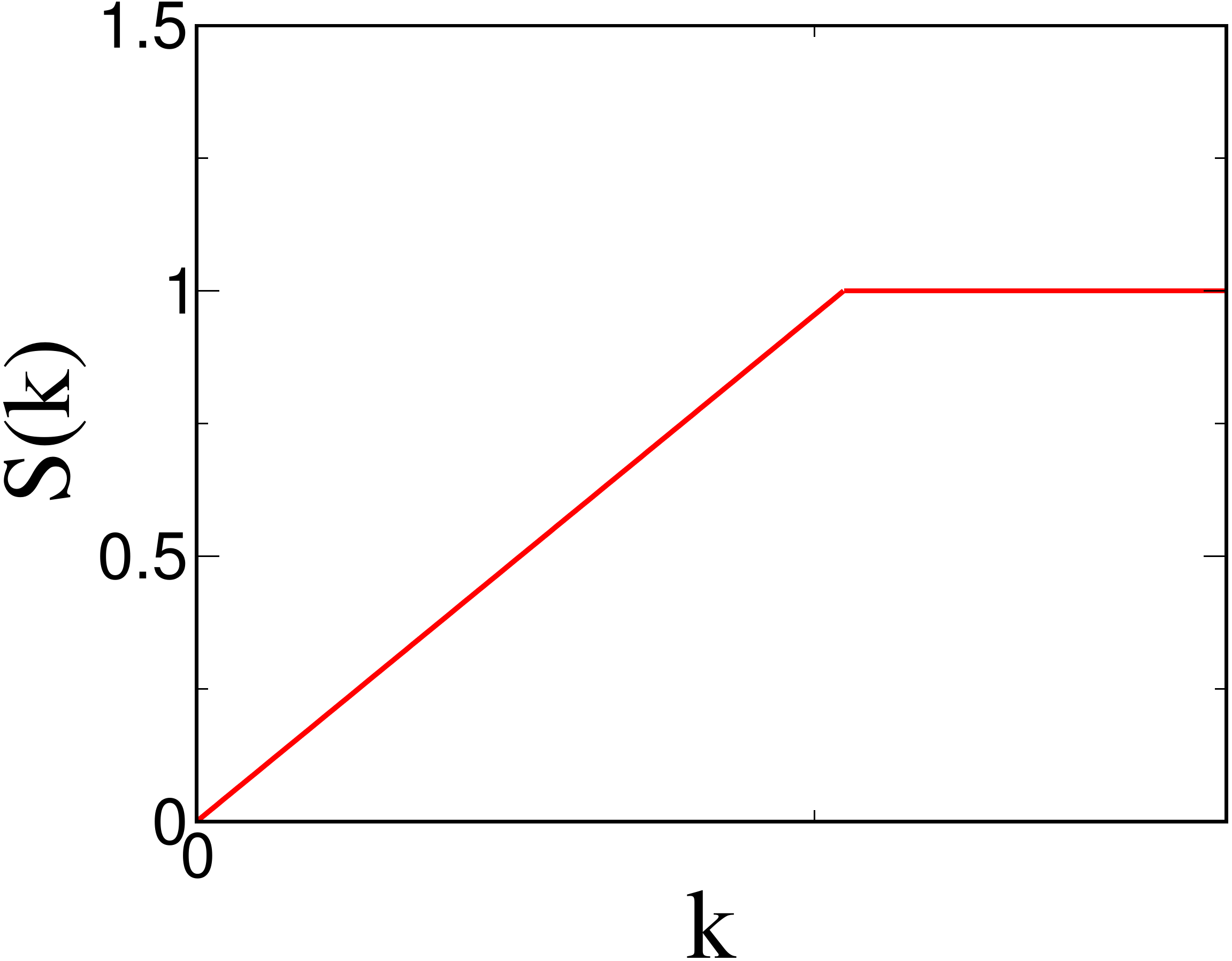}
}
\caption{Illustrative examples of structure factors for ordered and disordered point patterns in one-dimensional Euclidean space:
(a) Spatially uncorrelated (Poisson distribution or ideal gas); (b) Liquid; (c) Crystal (integer lattice); (d) Quasicrystal (Fibonacci chain);
(e) Limit-periodic (period-doubling chain); (f) Nontrivial zeros of the Riemann zeta function. Whereas the cases (a) and (b) are examples
of nonhyperuniform systems, the remaining cases represent hyperuniform examples.}
\label{examples}
\end{figure*}

X-ray and neutron scattering  techniques provide powerful ways to probe the structure of matter \cite{Chaik95}.
The observed scattering intensity in Fourier (reciprocal) space is encoded in the structure factor
$S({\bf k})$, where $\bf k$ is the wave vector.
A fundamental problem in condensed matter theory and statistical physics is the determination
of the class of ordered many-body systems with pure point diffraction or scattering patterns, i.e.,
those described by a  structure factor involving only a set of Dirac delta functions (Bragg peaks):
\begin{equation}
S({\bf k})= \sum_{j=1} a_j \delta({\bf k} - {\bf k}_j),
\label{Bragg}
\end{equation}
where the $a_j$ ($j=1,2,\ldots$) are positive weights.
It is well-known that crystals (periodic point patterns),
which have scattering patterns that are quite different from disordered systems (e.g., gases
and liquids) with continuous spectra,   fall in this class; see 
Fig. \ref{examples} for illustrative examples.  It came as a great surprise 
in the early 1980's that a family of noncrystalline (aperiodic) states of matter, called ``quasicrystals" \cite{Le84},
also have pure point diffraction, but with a twist, namely, they densely fill reciprocal space 
exhibiting symmetries that would be prohibited for crystals. A less familiar phylum of point patterns 
obeying (\ref{Bragg}) are limit-periodic systems \cite{Ba11,So11}. These are deterministic point patterns that consist of a
union of an infinite set of distinct periodic structures, 
and hence are also characterized by dense Bragg peaks. What differentiates limit-periodic systems from quasicrystals is that the ratio between any two 
peak locations is rational.

Where in this zoology of particle systems does one place the prime numbers? These are the numbers such as 2, 3, 5, 7, 11, 13 ... 163 ... 691 ... $2^{74,207,281}-1 \ldots$ having no proper factors, viewed as a one-dimensional point pattern. In the present paper, we show that the primes in  judiciously chosen intervals have dense Bragg peaks, similar to a limit-periodic system,
and hence satisfy (\ref{Bragg}). This is in astonishing contrast to the general understanding of primes
as pseudo-random numbers. 
The apparent difficulty of factoring large numbers into primes is basic to contemporary cryptography, and the lack of any obvious pattern is nicely summarized in a famous quotation attributed to R. C. Vaughan: ``{\it It is evident that the primes are randomly distributed but, unfortunately, we don't know what `random' means.}'' \cite{Gr95} In short intervals, say from a large number $X$ to $X+\ln(X)$, Gallagher \cite{gallagher1976distribution} proved that the primes have a pseudo-random spatial distribution with gaps following a Poisson distribution. In the present paper, we study longer intervals, such as $X$ to $2X$, and find that the primes have multiscale order characterized by dense Bragg peaks. Thus they obey (\ref{Bragg}), but are distinguished from both quasicrystals and limit-periodic systems in ways that we detail below. \\


Much of the modern understanding of prime numbers is based on a fundamental insight of Riemann \cite{Ri59}. He introduced what we now call the Riemann zeta function $\zeta(s)$, for a complex variable $s$, and indicated an explicit formula relating the primes to its zeros. Our original motivation to study the scattering patterns of the primes is the fact that the nontrivial zeros \cite{Note2} of the Riemann zeta function have an exotic hidden order on large length scales called {\it hyperuniformity}. A hyperuniform point configuration is one in which $S(k)$ tends  to zero as the wavenumber $k$ tends to zero  or, equivalently,  one in which
the local number variance $\sigma^2(R)$ associated within a spherical window of radius $R$
grows more slowly than $R$ in the large-$R$ limit ~\cite{To03a}. All perfect crystals and quasicrystals are hyperuniform,
but typical disordered many-particle systems, including gases, liquids, and glasses, are not. Disordered hyperuniform many-particle systems are exotic states of
amorphous matter that have attracted considerable recent attention in physics and materials science
because of their novel structural and physical properties \cite{Do05d,To08c,Ba08,Fl09b,Za11a, Ji14,To15,Ja15,Gh16,He17b,Ri17}.
According to the celebrated {\it Riemann hypothesis}, the nontrivial zeros of the zeta function lie along the {\it critical line} $s=1/2 + it$ with $t\in\mathbb{R}$ in the complex plane and thus form a one-dimensional point process. 
A resolution of this hypothesis is widely considered to be one of the most important open problem in pure mathematics \cite{Bo00}. 
Montgomery \cite{Mon73} advanced the conjecture that the pair correlation function $g_2(r)$ of the (normalized) zeros takes on the simple form $1-\sin^2(\pi r)/(\pi r)^2$. Remarkably, this exactly matches the pair correlation function of the eigenvalues of certain random Hermitian matrices \cite{Dy62a,Me91,Rud96}. The corresponding structure factor $S(k)$ tends to zero linearly in $k$ in the limit $k\to 0$, as shown in Fig. \ref{examples}f.  This means that the
Riemann zeros are disordered but hyperuniform~\cite{To08c}. In his famous essay entitled ``Birds and Frogs" \cite{Dy09}, Dyson  suggested an approach to the Riemann Hypothesis where one first classifies all one-dimensional quasicrystals and then shows that one such quasicrystal  corresponds to the non-trivial Riemann zeros.



An important aspect of the distribution of prime numbers is that larger ones become increasingly sparse. The drop-off is gradual enough that, in Gallagher's regime of short intervals or even for the longer intervals considered here, the density of prime numbers can be treated as constant \cite{To17b}. According to the {\it prime number theorem} \cite{hadamard1896distribution}, the prime counting function $\pi(x)$, which gives the number of primes less than $x$, in the large-$x$ asymptotic limit is given by
\begin{equation}
\pi(x) \sim \frac{x}{\ln(x)} \qquad (x \rightarrow \infty).
\end{equation}
One can interpret this as indicating that the probability that a randomly selected integer less than a sufficiently large $x$ is prime is inversely proportional to the number of digits of $x$. 
This implies a position-dependent number density $\rho(x) \sim 1/\ln(x)$. Thus the primes constitute a {\it statistically inhomogeneous} set of points in large intervals, 
becoming sparser as $x$ increases. This means one must be careful in choosing the interval over which the primes are sampled. This observation is crucial to the remarkable properties of the primes that we report here.
 
The hyperuniformity of the Riemann zeros led us to seek intervals in the primes in which they  might be regarded as a hyperuniform point pattern. In a concurrent numerical study \cite{Zh17}, we  examined the structure factor $S(k)$ for primes in an interval $[M,M+L]$ with $M$ large (say, $10^{10}$) and $L/ M$ a small positive number. These simulations strongly suggest that the structure factor in such finite intervals exhibits many well-defined Bragg-like peaks dramatically overwhelming a small ``diffuse" contribution, indicating that the primes are more ordered than previously known.
 
Motivated by this numerical study, here we apply the tools of statistical physics  to understand the
nature of the primes as a point process by quantifying the   structure factor, pair correlation function,
local number variance,  and the $\tau$ order metric in various intervals. 
Our main results are obtained for the  interval $M \leq p \leq M + L$ with 
$M$ very large and the ratio $L/M$ held constant. This enables us to treat the primes as a homogeneous point pattern. We also consider appreciably larger and smaller intervals for purposes of comparison.
We prove that the primes are characterized by unanticipated multiscale order; see Ref. \cite{To17b} for details.
Specifically, an analytical formula that we derive for their limiting structure factor $S(k)$ has dense Bragg peaks, as in the case of quasicrystals \cite{Le84}. Unlike quasicrystals, however, the prime peaks occur at certain rational multiples of $\pi$, which is similar to limit-periodic systems \cite{Ba11}. But the primes show an erratic pattern of occupied
and unoccupied sites, very different from the predictable and orderly patterns of standard limit--periodic systems. Hence, the primes are the first example of a point pattern that is {\it effectively} limit-periodic. 

Our analysis is rooted in the {\it circle method} of Hardy-Littlewood \cite{Ha23}, in particular their conjecture on prime $k$-tuples, but we emphasize the perspective of statistical physics and the new consequences that arise
in the limit of infinite system size. Our analytical formula (\ref{g2eq}) expresses the pair correlation function $g_2$, including the density of twin primes, as an infinite sum, whereas the celebrated Hardy-Littlewood representation was originally presented as a product over primes. 
Using a scalar order metric $\tau$ numerically calculated from $S(k)$, we identify a transition between the order exhibited when $L$ is comparable to $M$ and the uncorrelated behavior when $L$ is only logarithmic in $M$. Our formulation also yields an algorithm that enables one to predict (reconstruct) primes  with high accuracy.

\section*{Results}

We consider the primes in the interval $[M,M+L]$ to be a special `lattice-gas' model: the primes and odd composite integers are  
``occupied'' and ``unoccupied" sites, respectively, on an integer lattice of spacing 2 that contains all of the positive odd integers. We study the pair statistics between primes in such intervals.
If $L$ is much larger than $M$, the density $1/\ln(n)$ drops off appreciably as $n$ ranges from $M$ to $M+L$, and then 
we show that the system is diametrically the opposite of hyperuniform. On the other hand, if the interval is small such that $L \sim \ln(M)$, one enters Gallagher's regime in which the primes are Poisson distributed.

One of our main analytical result is formula (\ref{SkPeak}) for the structure factor of the primes, valid in the regime $M \rightarrow \infty$ with $L/M$ converging to a fixed positive value. We denote this ratio of $L$ to $M$ by $\beta$. Our limiting form for $S(k)$, given by Eq. (\ref{S-p}),  is valid for any positive value of $\beta > 0$, as long as $\beta$ does not vary with $M$. These results lead
to several significant consequences, which we describe below.

\begin{figure}[bt]
\includegraphics[width=0.4\textwidth,clip=]{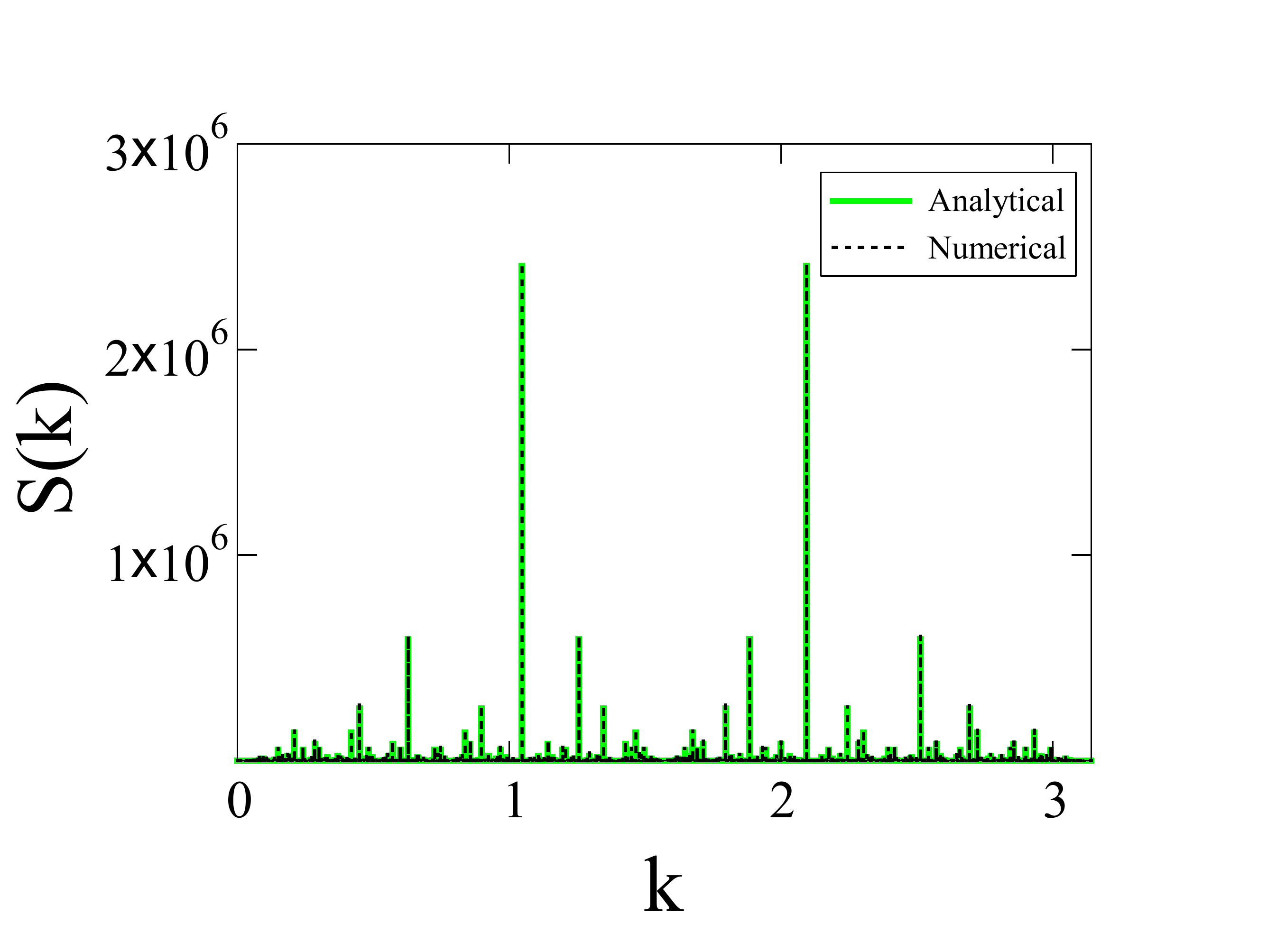}
\includegraphics[width=0.4\textwidth,clip=]{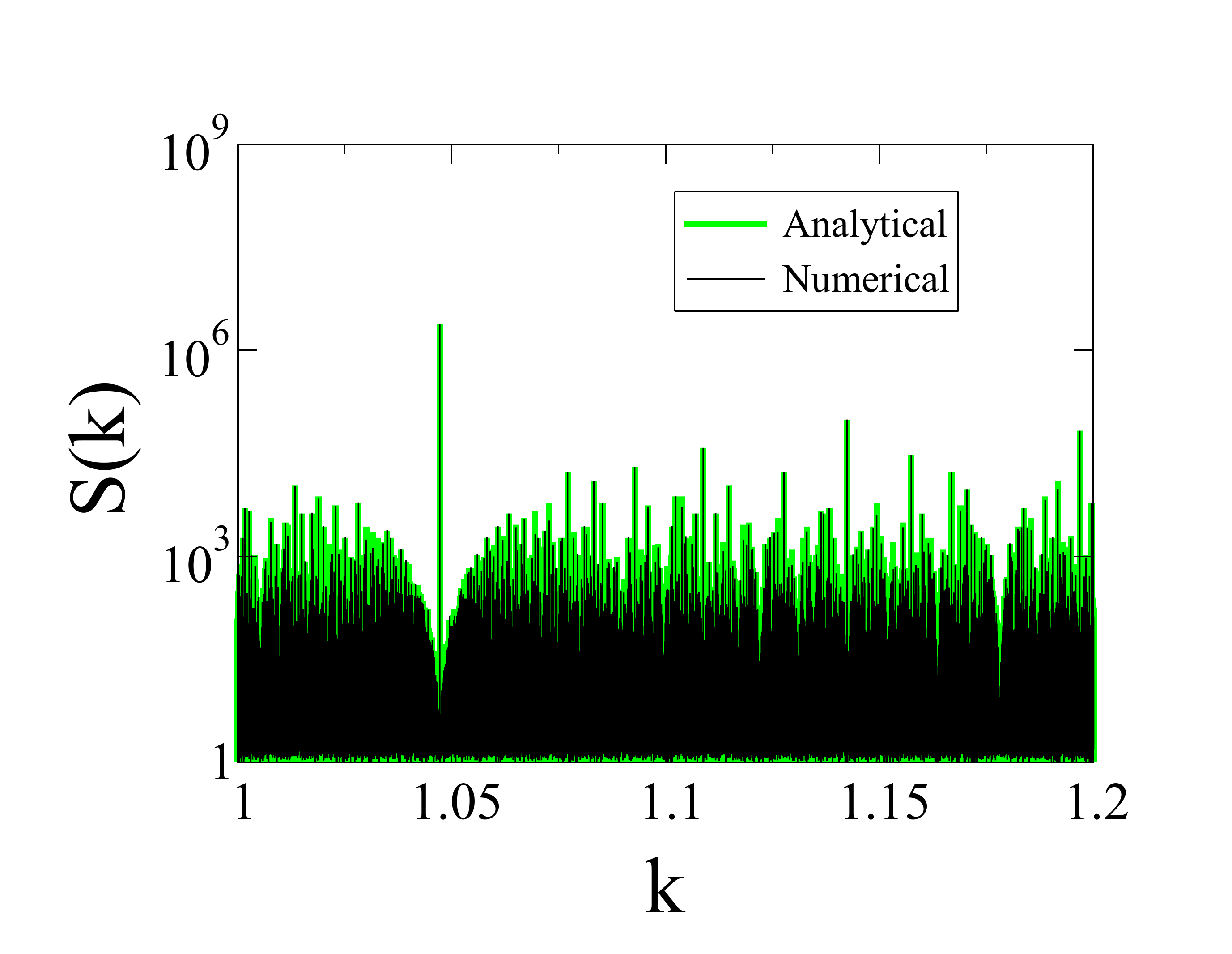}
\includegraphics[width=0.35\textwidth,clip=]{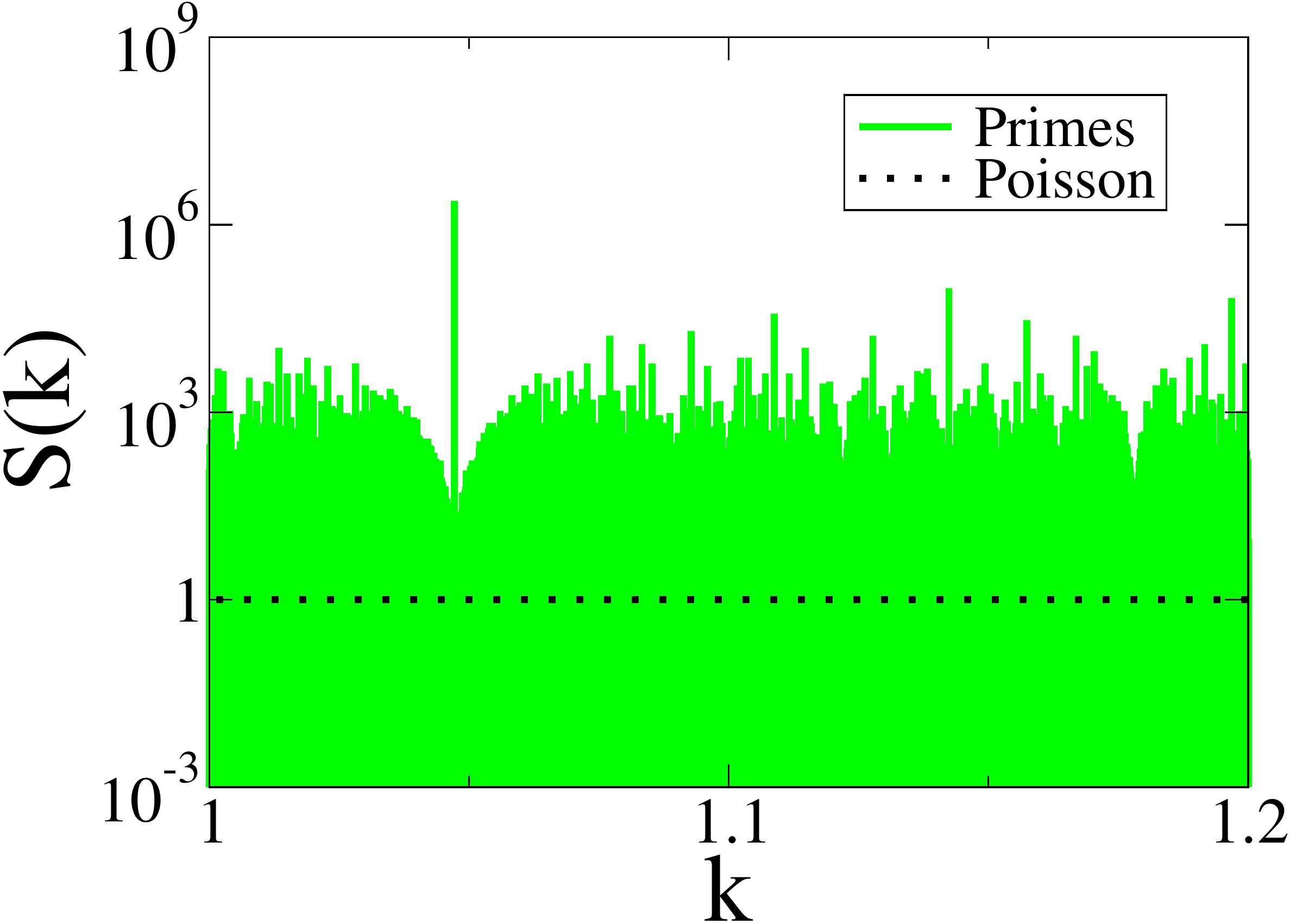}
\caption{Top: $S(k)$ for the primes as a function of $k$ (in units of the integer lattice
spacing), as predicted from formula (\ref{SkPeak}) for $M=\ee{10}+1$, $L=2.23\times\ee{8}$ and $n_{max} =100 \ln(M)$
is in excellent agreement with the corresponding numerically computed structure factor obtained in Ref. \cite{Zh17}.
Note the many Bragg peaks of various heights
with a self-similar pattern. Middle: Same as the top panel, but 
at a smaller horizontal scale and log vertical scale to reveal the dense peak structure. Bottom: The prediction from (\ref{SkPeak}) but where the horizontal scale is 
appreciably smaller than the top panel to again reveal the dense peak structure and its stark contrast with the uncorrelated (Poisson) lattice gas.}
\label{fig:Sk}
\end{figure}

We study various prime intervals, 
but we show that when $L \sim \beta M$, the major contribution to the structure factor,
$S(k)$ is a set of dense Bragg peaks that are located at certain rational values of 
$k/\pi$ with heights given by 
\begin{equation}
S(\pi m/n) \sim \frac{N}{\phi(2n)^2} \mu(2n)^2,
\label{SkPeak}
\end{equation}
where $N$ is the number of primes in the interval from $M$ to $M+L$, $m \le n$ and $n$ are co-prime integers  (share no common divisors, except 1), 
$\phi(n)$ is Euler's totient function \cite{Te95}, which counts the positive integers up to a given integer 
$n$ that are co-prime to $n$, and $\mu(n)$ is the M\"{o}bius function \cite{Te95} so that $\mu^2(2n)$
is one whenever $2n$ is square-free and zero otherwise. Notice that as $n$ grows, the size of the peak shrinks because of the denominator $\phi(n)^2$.
Figure \ref{fig:Sk} depicts the structure factor
of the primes obtained from (\ref{SkPeak}) at different different horizontal scales, where $M=\ee{10}+1$, $L=2.23\times\ee{8}$  and $n$ is truncated at $n_{\max}=100 \ln(M)$.  
This is in excellent agreement with the corresponding numerically computed structure factor of the
actual primes configuration in this interval (top and middle panels of Fig. \ref{fig:Sk}).
The structure factor contains many well-defined Bragg-like peaks of various intensities
characterized by a type of self-similarity. This self-similarity follows from the fact that $\phi(n_1 n_2) = \phi(n_1)\phi(n_2)$ for relatively prime $n_1$ and $n_2$, so that rescaling preserves the relative heights of the peaks given by Eq. (\ref{SkPeak}).  The bottom
part of the figure  zooms in on small wavenumbers and compares  $S(k)$ 
to that of an uncorrelated lattice gas. This definitively demonstrates that the 
primes in this interval exhibit order across multiple length scales, making 
them substantially more ordered than the uncorrelated lattice gas found by Gallagher for shorter intervals of primes.

\begin{figure*}[bthp]
\includegraphics[width=1\textwidth]{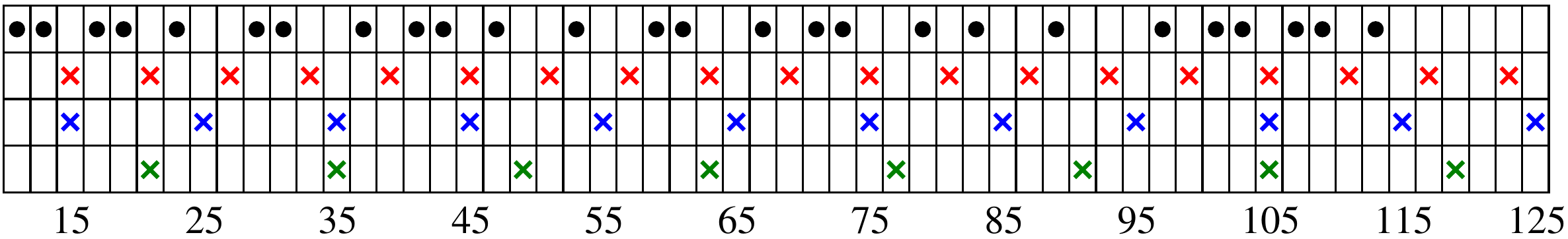}
\caption{Illustration of the superposition of effective multiple periodicities in the primes.
We take the primes to be ``occupied'' sites (black dots) on an integer lattice  of spacing 2 that contains all of the positive odd integers.
The crosses indicate sites that cannot be occupied because of a certain periodicity $2n$ ($n$ sites on the odd integer lattice), where $n$
is a square-free odd number. For example, the peak at $\pi/3$ with $n=3$ and $m=1$ corresponds to remainders when dividing by 6. 
A prime must leave a remainder of 1 or 5 or else it would be divisible by 2 or 3. The lattice in the figure has a spacing of 2, so these allowed sites appear with a period of 3 instead of 6.  The forbidden sites appear as red crosses. The other two sites may or may not be prime, and if one averages over many periods in a sufficiently large interval,
 each of them will  have an equal occupation probability 
(due to Dirichlet's theorem on arithmetic progressions).
The overall effect of this equi-distribution of occupied sites is an effective periodicity of 6.
Similarly, the primes  show an effective periodicity of 10 (blue crosses), 14 (green crosses), and even larger periods (not shown in the
figure). The superposition of all of the effective periodicities leads to a pattern
of dense Bragg peaks located at $m\pi/n$ (where $m$ are the positive integers and $n$ is odd and square-free), reminiscent of a limit-periodic system even though each local period is subject to erratic disruptions. These peaks are illustrated in Fig.~\ref{fig:Sk}. However, if the interval is too small or too large, then the effective periodicity would not be seen. It is a distinctive feature of primes in intervals from $M$ to $M+L$ with $M$ and $L$ large numbers of comparable magnitude.}
\label{cartoon}
\end{figure*}

{\it ``Effective" Limit-Periodicity.} The proof of (\ref{SkPeak}) is given  in Ref. \cite{To17b}. 
It is based on grouping the terms in the sum defining $S(\pi m/n)$ according to their remainder after division by $2n$. A fundamental heuristic about prime numbers is that each of the allowed remainders occurs roughly equally often, so that the primes are evenly distributed in arithmetic progressions. This breaks down if the modulus $n$ is too large relative to the primes under consideration. Increasingly precise versions of this statement have been established rigorously, starting from Dirichlet's theorem that there are infinitely many primes for each remainder to a fixed modulus. A strong interpretation conjectured by Elliott-Halberstam plays a role in some of the work of the Polymath project pushing progress on gaps between primes to the limit. The key calculation underlying our proof is that the roughly even distribution across all possible remainders causes constructive or destructive interference in the sum $S(\pi m/n)$, depending on the fraction $m/n$, and leads to our proof. Referring to (\ref{SkPeak}), note that if  $n$ is even or if $n$ has a repeated factor, 
then $\mu(2n) = 0$ so $S(\pi m/n)$ vanishes up to the accuracy in comparing the number of primes in different progressions. If $n$ is odd and square-free, then the structure factor has a peak of size $N/\phi(n)^2$ (since, $n$ being odd, $\phi(2n)=\phi(n)$). 
This explains the peaks observed numerically \cite{Zh17} at, for example, $S(\pi/3)$.  A value on the order of $N$ should indeed be viewed as a peak since $S(k)$ is a sum of length $N$ and ignoring all cancellation shows that $|S(k)| \leq N$. 
Thus the largest values of $S(k)$ are the peaks when $k/\pi$ is a rational number  with odd, square-free denominator. Taken together, 
these locations correspond to effective periodicities, as illustrated in Fig. \ref{cartoon}.

In the limit of infinite system size, the peaks will become Dirac delta functions
at rational numbers with odd, square-free denominators, and the discrete formula (\ref{SkPeak})
(scaled by $2\pi \rho$) tends to 
\begin{equation}
\lim_{M \rightarrow \infty} \frac{S(k)}{2\pi \rho} = {\sum_{n}}^{\flat} {\sum_{m}}^{ \times} \frac{1}{\phi(n)^2} \delta\left(k - \frac{m\pi}{n}\right),
\label{S-p}
\end{equation}
where the symbol $\flat$ is meant to indicate that the sum over $n$ only involves odd, square-free values of $n$
(excluding 1 to eliminate forward scattering)
and the symbol $\times$ indicates that $m$ and $n$ have no common factor \cite{Note1}.  
This implies that the ``diffuse" part observed numerically in Ref. \cite{Zh17} vanishes in the infinite-system-size limit.
Hence, the primes become effectively limit-periodic, despite the variable pattern of occupied sites,
as proved in Ref. \cite{To17b} and illustrated in Fig. \ref{cartoon}. Such multiscale order in the primes
appears to be a new discovery.

This is to be distinguished from a standard limit-periodic system, which  is a deterministic
point process characterized by dense Bragg peaks at rational multiples of $\pi$. A prototypical
example is the {\it period-doubling chain}  \cite{Ba11}. In this model, there are sites of two types, $a$ and $b$, forming a point pattern on the integer lattice defined by the following iterative substitution rule, initialized with a single site $a$: 
$a \rightarrow ab$ and $b \rightarrow aa$ \cite{Ba11}.
 The locations of the $b$'s are given by a superposition of arithmetic progressions 
$2+4j$, $8+16j$, $32+64j$, with a factor of 4 from one to the next. Thus, the infinite-size limit is a union of periodic systems in 
which $S(k)$ consists of dense Bragg peaks at certain rational values $k/\pi$. The structure factor associated with the 
$a$'s (assuming unit lattice spacing) is given by
\begin{eqnarray}
S(k) &=& \frac{4\pi}{3}\Bigg[\sum_{m=1}^{\infty} \delta(k-2\pi m) \nonumber \\
&+& \sum_{n=1}^{\infty} \sum_{m=1}^{\infty} 2^{-2n} \delta\left(k-\frac{(2m-1)\pi}{2^{n-1}}\right) \Bigg].
\label{S-doubling}
\end{eqnarray}
Figure \ref{Bragg}e shows the structure factor for the period-doubling chain.

{\it Hyperuniformity.}--~ We now show that the effective limit-periodic form  (\ref{S-p}) of $S(k)$ implies
that the primes are  hyperuniform. The structure factor $S(k)$ is not a continuous function because there are dense Bragg peaks arbitrarily close to 0, so we 
do not have $S(k) \rightarrow 0$ as $k \rightarrow 0$ in the usual sense. We follow the practice of Ref. \cite{Og17} in such instances 
and pass to a cumulative version of the structure factor,
$Z(K)$, defined by 
\begin{equation}
Z(K) = 2 \int_0^K  S(k)  dk,
\label{Zofk}
\end{equation}
which is the {\it cumulative} intensity function within a ``sphere" of radius $K$
of the origin in reciprocal space. If  $Z(K)$ tends to 0 as a power $K^{\alpha+1}$, any positive power $\alpha > 0$ yields hyperuniformity and distinguishes the 
primes from a Poisson distribution of points with the same density.
Using relations (\ref{S-p}) and (\ref{Zofk}), we  find 
\begin{equation}
\lim_{M \rightarrow \infty} \frac{Z(K)}{2\pi \rho} = 2{\sum_{n}}^{\flat} {\sum_{m\pi/n < K}}^{\hspace{-0.1in}\times} \frac{1}{\phi(n)^2}.
\label{limit}
\end{equation}
Using (\ref{limit}) together with some results from analytic number theory \cite{To17b}, we can show that $Z(K) \sim K^2$ as $K\rightarrow 0$.

A one-dimensional hyperuniform point process is one in which $\sigma^2(R)$
grows more slowly than $R$ in the large-$R$
limit. Using formula (\ref{eqn:local-1}) that relates  $\sigma^2(R)$ to $Z(K)$, we find relation (\ref{limit}) implies
the primes  have a number variance $\sigma^2(R)$ that scales logarithmically with $R$ in the large-$R$ limit. This is precisely the
same growth rate exhibited by the Riemann zeros \cite{To08c}, but
as we will see, the latter are appreciably less ordered than the former.


{\it Transition Between Order and Disorder.}--~ A useful scalar quantity that is capable of capturing the
degree of translational order of a point process  in Euclidean spaces across length scales is the $\tau$ order metric  
\cite{To15}. Here we use the discrete-setting counterpart of this order metric  
for a lattice gas  in a fundamental cell of length $L$ under periodic boundary conditions \cite{Di18}:
\begin{eqnarray}
\tau &=& \frac{1}{N_s} \sum_{j=1}^{Ns-1}\left(S\left(\frac{j\pi}{N_s}\right)-(1-f)\right)^2,
\label{tau-discrete}
\end{eqnarray}
where $N_s$ is the number of lattice sites within the fundamental cell and $f$ is the occupation fraction.
In the case of an ensemble-averaged uncorrelated lattice gas, $S=1-f$  in the infinite-system-size limit
so that $\tau=0$. Thus, a deviation of $\tau$ from
zero measures translational order with respect to the fully uncorrelated case.
For lattice gases  characterized by Bragg peaks,
$\tau/\rho^2$ grows linearly with  $L$  for sufficiently large $L$:  
\begin{equation}
\tau/\rho^2 \sim c L,
\label{tau-L}
\end{equation}
where $c$ is dependent on the system. Based on this order metric,  the primes are substantially  more ordered than the uncorrelated lattice gas
and appreciably less ordered than an integer lattice, but similar in order to the period-doubling chain \cite{Ba11}.
For example, consider the integer lattice with spacing of occupied sites such that $f=0.1$, 
chosen to match the density of our system of primes. The lattice has $c=18$, much larger than the value $c=0.1674$ for primes, which are closer to the period-doubling chain ($c=0.1429$). 
In all of these ordered examples, $\tau$ grows with the system size $L$. 
This indicates multiscale order in the primes, absent from a case such as the Riemann zeros in  which, assuming Montgomery's pair 
correlation conjecture,  $\tau$ converges to 2/3 and in particular does not grow with $L$. \cite{To17b}.

\begin{figure}
\includegraphics[width=0.45\textwidth]{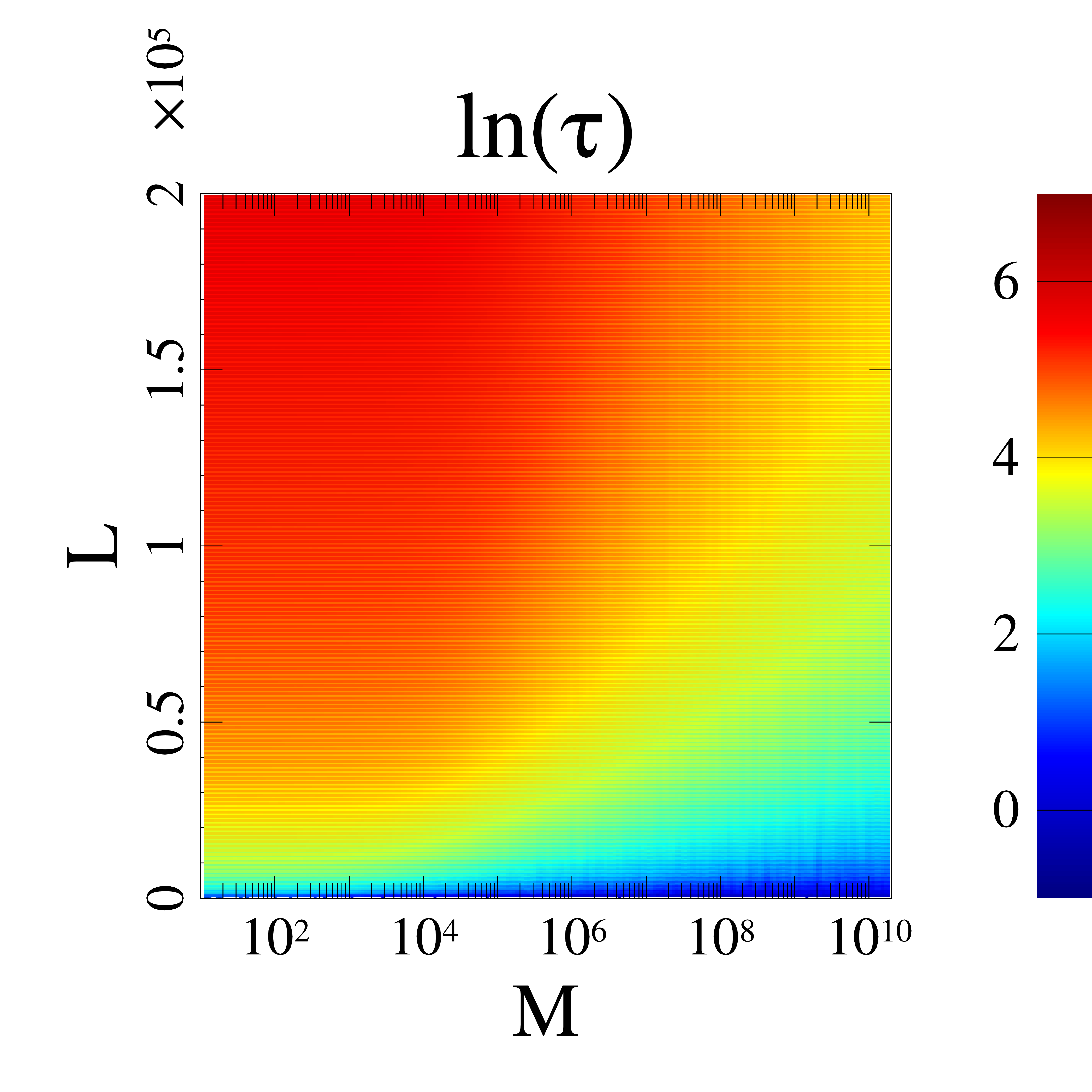}
\caption{Natural logarithm of the order metric $\tau$ of prime numbers for $10<M<2\times 10^{10}$ and $8<L<2\times 10^5$.
The ``warmer" colors in the upper left corner indicate systems with a larger value of $\tau$, and hence stronger order. The ``cooler" colors in the lower right indicate relatively disordered  systems comparable to an uncorrelated lattice gas. The interfaces separating these layers have a parabolic shape, which is explained by (\ref{tau-L}).}
\label{tau}
\end{figure}

Now we study the $\tau$ order metric for prime-number configurations with different $M$ and $L$. 
As derived from (\ref{tau-L}) and illustrated by Fig.~\ref{tau}, the constant-$\tau$ level curves have the form $L \sim \ln^2(M)$, i.e., level curves appear
as quadratic curves in the log plot.
For an uncorrelated lattice gas for any  $L$ and $\rho$, $\tau$ is very small.
Thus, $L \sim \ln^2(M)$  is the boundary between regions where primes can be considered to be uncorrelated versus 
correlated. For the uncorrelated regime in which Gallagher's results
apply, $L\sim\ln(M)$, and $\tau$  diminishes as $M$ increases. As $L$ increases, prime-number configurations move from the uncorrelated regime ($\tau\sim 1$, $L \le \ln^2(M)$) to the limit-periodic
regime we studied in this paper ($\tau\sim L$, $L \propto M$), and then to the inhomogeneous and nonhyperuniform regime, where the density gradient is no longer
negligible (e.g., if $L \sim M^2$).

{\it Recovery of Hardy-Littlewood Conjecture.}--~ 
We can get the pair correlation function $g_2(r)$ of the primes by performing the inverse Fourier transform of $S(k)-1 \equiv \rho {\tilde h}(k)$ using (\ref{S-p}),
where ${\tilde h}(k)$ is the Fourier transform of $h(r)\equiv g_2(r)-1$ for $r\neq 0$:
\begin{equation}
g_2(r)=1+{\sum_{n}}^{\flat} \frac{1}{\phi^2(n)}{\sum_{m}}^{\times} \exp(rm\pi i/n).
\label{g2eq}
\end{equation}
The expression  given in Eq. (\ref{g2eq})  for distinct values of $r= 2, 4, 6,\ldots$ is a different representation of the 
constants in  the famous Hardy-Littlewood  $k$-tuple conjecture (Theorem X, p.61 in \cite{Ha23}) for the special case $k=2$. The Hardy-Littlewood conjecture is beyond doubt in the mathematical community, and yet far from a rigorous proof. Spectacular progress towards it was made recently by Maynard, Zhang, Tao, and the massive online collaboration Polymath 8 \cite{zhang_2014,Poly14,maynard_2015}. In particular, Maynard's Theorem 1.2 from \cite{maynard_2015} shows that there must be infinitely many prime shifts for a positive proportion of $k$-tuples.

Summation of (\ref{g2eq}) for $r=2,4,6,8$ and 10 with a cut-off $n<10000$ yields
predictions that are in agreement with established values, as shown in Table \ref{Hardy}.
Indeed, we can prove  \cite{To17b} that Eq.~(\ref{g2eq}) is equivalent to Hardy and Littlewood's original expression \cite{Ha23}.
This adds to the validity of our effective limit-periodic form of the structure factor of the primes,
which has heretofore not been identified.

\begin{table}[H]
\centering
\caption{Hardy-Littlewood constants for $k=2$ 
calculated with our formula (\ref{g2eq}) using $n_{\mbox{max}}=10000$ for $r=2,4,6,8$ and 10 compared to their accurate values 
\cite{Ha23}. The case $r=2$ corresponds to the ``twin" primes. Notice that $g_2(2)=g_2(4)=g_2(8)$, which has a simple explanation in terms of the Hardy-Littlewood product over primes. This product includes a special contribution from primes dividing $r$, and the only such prime is $2$ when $r$ is a power of 2.}
\label{Hardy}
\begin{tabular}{c|c|c}
r  & Prediction of (\ref{g2eq}) & Ref. \cite{Ha23} \\ \hline
2  & 0.660161536    & 0.660161816               \\
4  & 0.660161536    & 0.660161816               \\
6  & 1.320323071    & 1.320323632               \\
8  & 0.660161536    & 0.660161816               \\
10 & 0.880215710    & 0.880215754              
\end{tabular}
\end{table}
\vspace{-0.2in}

\begin{figure}
\includegraphics[width=0.45\textwidth]{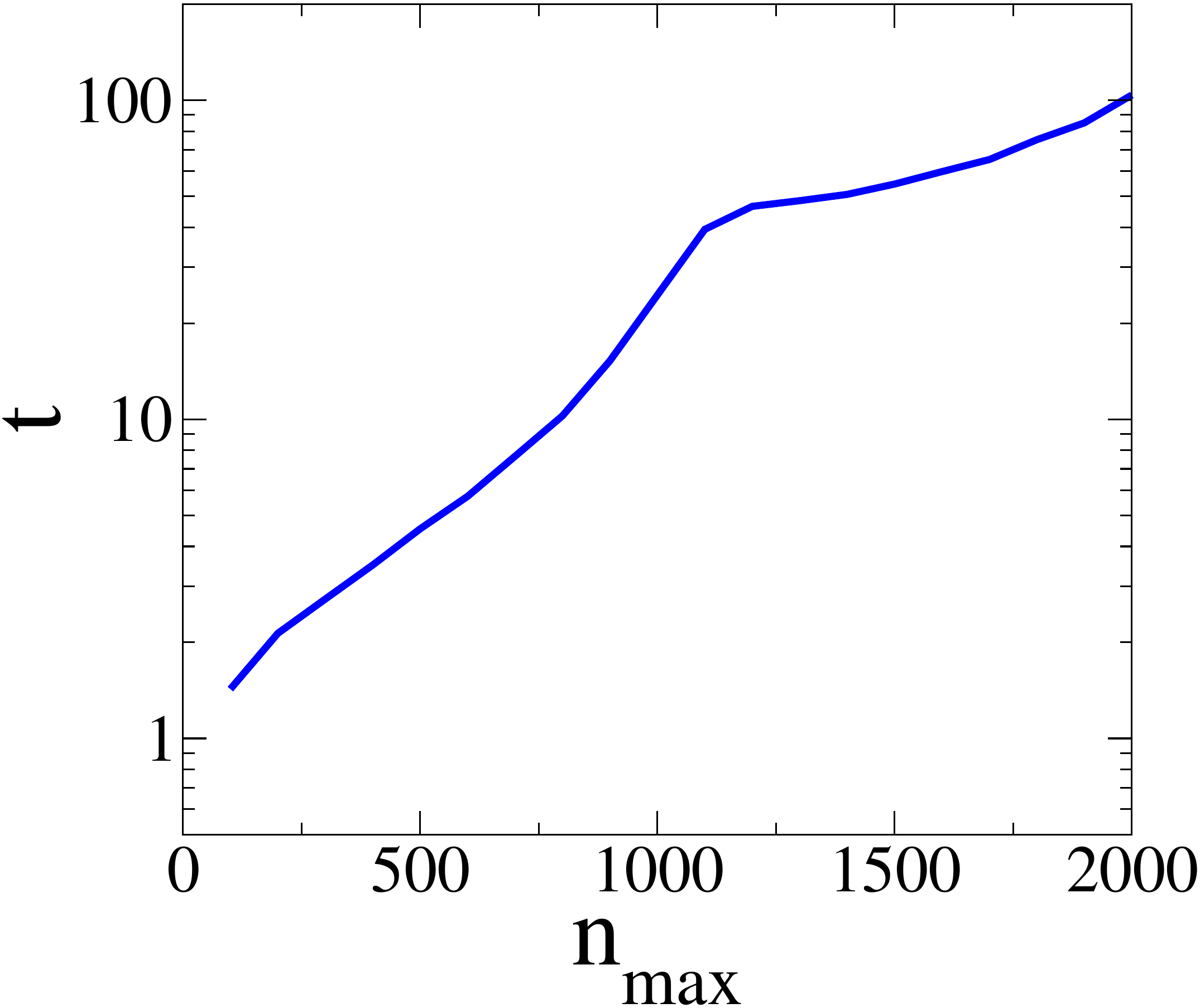}
\caption{Prime prediction accuracy parameter $t$  versus $n_{max}$ for  $M= 10^{6}$.}
\label{recon}
\end{figure}

{\it Reconstruction of the Prime Numbers}--~Note that we not only have an analytical formula for $S(k)$, but also for the
complex density variable ${\tilde \eta}(k)$, defined by (\ref{etak}), which includes phase information.  
This analytical expression for ${\tilde \eta}(k)$ of the primes enables us to reconstruct, in principle,  a prime-number configuration within an arbitrary interval $[M, M+L]$ by obtaining the inverse Fourier transform of ${\tilde \n}(k)$.
In practice, one is computationally limited by the fact that one can only include
a finite number of peaks for which $n<n_{max}$; see Ref. \cite{To17b} for details.
This leads to an algorithm to  reconstruct primes in a dyadic interval with high accuracy provided
that $n_{max}$ is sufficiently large and $M$ is not too large. A measure of the accuracy
of the reconstruction algorithm is given by the ratio $t=N_c/N_i$, where
$N_c$ and $N_i$ are the number of correctly predicted primes and incorrectly predicted 
primes (composite odd numbers), respectively. Figure \ref{recon} shows that
the reconstruction procedure for $M=10^6$ becomes highly accurate as the number of Bragg
peaks incorporated, as measured by $n_{max}$, increases.

\section*{Discussion}

In summary, by focusing on the scattering characteristics of the primes in certain sufficiently large intervals,
we have discovered that prime configurations  are hyperuniform and characterized by 
an unexpected order across length  scales.  In particular, they provide
the first example of an  effectively limit-periodic point process, a hallmark of
which are dense Bragg peaks in the structure factor. The discovery of this hidden 
multiscale order in the primes is in contradistinction 
to their traditional treatment as pseudo-random numbers. 

Effective limit-periodic systems represent a new class of many-particle systems
with pure point diffraction patterns that deserve future investigation in physics, apart
from their connection to the primes. For example, the formulation  of other theoretical 
structural models of effectively limit-periodic point processes in one and higher
dimensions and the study of their physical properties are exciting areas for further exploration.

\section*{Methods}

As noted earlier, we consider the primes in the interval $[M,M+L]$ as lattice-gas in which the primes
occupy a subset of the odd integers.  A particular lattice-gas configuration under periodic boundary conditions
is characterized by the {\it local} density:
\begin{equation}
\n({\bf r})=\sum_{i=1}^N \delta({\bf r} -{\bf x}_i),
\label{n}
\end{equation}
where $N$ is the number of primes in the interval.
The Fourier transform of the local density, called the {\it complex collective density variable} ${\tilde \n}({\bf k })$, is given by
\begin{equation}
{\tilde \n}({\bf k }) = \sum_{j=1}^{N} \exp(-i{\bf k \cdot r}_j).
\label{etak}
\end{equation}
The corresponding structure factor is given by
\begin{equation}
S(k) \equiv  \frac{|{\tilde \n}({\bf k })|^2}{N}
\end{equation}
where the wavenumber $k$ ranges from zero to $\pi$, extended periodically. Ultimately, we pass to the limit $N \to \infty$.


The local number variance $\sigma^2(R)$ associated
within  an interval (window) of length $2R$ for
a one-dimensional homogeneous point process \cite{To03a}
depends on an integral involving the structure factor $S({\bf k})$:
\begin{eqnarray}
\sigma^2(R)&=&
\frac{2\rho R}{\pi} \int_0^{\infty} S(k)
{\tilde \alpha}_2(k;R) dk,
\label{local}
\end{eqnarray}
where ${\tilde \alpha}_2(k;R)= 2 \sin^2(k R)/(kR)$. After integrating by parts, the second line of (\ref{local}) leads to an alternative representation of the number variance \cite{Og17}:
\begin{equation}
\sigma^2(R)=
-\frac{\rho R}{(\pi)} \int_0^\infty  Z(k) 
\frac{\partial {\tilde \alpha}_2(k;R)}{\partial k} dk,
\label{eqn:local-1}
\end{equation}
where $Z(K)$ is defined by (\ref{Zofk}). The quantity  $Z(k)$ has advantages over $S(k)$ in the characterization of quasicrystals and other point processes with dense Bragg peaks \cite{Og17}. This is the formula that we employ to determine the variance for the primes.
\bigskip

{\noindent{\bf Acknowledgments:} We are grateful to Peter Sarnak for valuable discussions.
This work was supported in part  by the National Science Foundation under Award No. DMR-1714722.
de Courcy-Ireland was supported by the Natural Sciences and Engineering Research Council of Canada.


%

\vspace{0.2in}


\end{document}